\documentclass[aps,prb,twocolumn]{revtex4}
\usepackage{graphics}
\begin{document}
\hoffset -0.8cm
\large

\title{ 
\Huge \bf How to determine model Hamiltonians for strongly
correlated materials \vspace*{2cm}}

\author{\LARGE Marie-Bernadette LEPETIT} 

\affiliation{\em \LARGE Laboratoire de Physique Quantique \\ 
Unit\'e Mixte de Recherche 5626 du CNRS \\ 
IRSAMC \\ 
Universit\'e Paul Sabatier \\
118 route de  Narbonne \\ 
F-31062 Toulouse \\ 
France}

\maketitle
\newpage



\section*{\underline{\Large Abstract} }
The present paper reviews recent achievements on the ab initio
determination of effective model Hamiltonians aimed at the description
of strongly correlated materials. These models (Heisenberg, $t-J$,
extended Hubbard, Kondo, etc) are crucial to solid state physicists
for the description and understanding of the electronic remarkable
properties such as magnetic orders, photo-induced magnetism, transport
properties, high $t_c$ super conductivity, etc.  We will see how the
association and the control of embedding techniques, fragment
spectroscopy, effective or intermediate Hamiltonian theories,
provides a systematic and well controlled method for the determination
of trustworthy models Hamiltonians.

\section*{\underline{\Large 1. Introduction} }
\label{sec:intro}

Since the synthesis of the first organic conductors in the late 70s
with the $\rm TCNQ$-based compounds~\cite{TCNQ} and later the
Bechgaard salts~\cite{Bech}, the chemists have synthesized a large
number of materials presenting new and remarkable electronic
properties. Besides the organic and organo-metallic metals, to cite
only a few among the most attractive systems of the last decades, one
can remember the spin chains~\cite{spch} and ladders~\cite{spld}, the
copper-oxide high $T_c$ super-conductors~\cite{htc}, manganites with
giant magneto-resistance~\cite{mangn}, photo-induced
magnets~\cite{verd}, etc...  These materials span a very large variety
of structural arrangements, phases and physical properties. Indeed one
can find among them molecular crystals, transition metals oxides,
Prussian blue analogs, etc. They can be isotropic as well as present
large anisotropies both in their properties and structural
arrangement, from the cubic, 3D, isotropic photo-induced magnets, to
the 2D, layered copper oxides or manganites, up to the quasi-1D stacks
of the Bechgaard salts or oxides spin chains and ladders. These
compounds present a wide variety of phases going from metals and
super-conductors to semi-conductors, Peierls or spin-Peierls
insulators, Mott insulators, charge or spin density wave insulators,
they may present long range order such as anti-ferromagnetic ordering,
have exponentially vanishing correlation functions, etc.

In view of this very large spectra of chemical compositions,
structural arrangements and physical properties one may wonder what
all these compounds have in common. In fact all these materials have
in common an essential characteristic~: the electrons responsible for
their low energy physics are few (per unit cell) and localized both
spatially and energetically. 
The spatial localisation of these, from now on called, {\em active}
electrons induces a strong correlation between their relative
positions ($\rho_2\left( \vec r_1,\vec r_2\,;\,\vec r^\prime_1,\vec
r^\prime_2\right) \ne \rho_1\left( \vec r_1 \,;\, \vec r^\prime_1
\right) \rho_1\left( \vec r_2 \,;\, \vec r^\prime_2 \right) -
\rho_1\left( \vec r_1 \,;\, \vec r^\prime_2 \right) \rho_1\left( \vec
r_2 \,;\, \vec r^\prime_1 \right)$, where $\rho_2$ is the two electrons
density matrix and $\rho_1$ is the one electron density matrix) and
the impossibility to describe their movement ---~even in a qualitative
manner or at the zeroth order~--- by an independent electrons
representation.  These systems belong to the strongly correlated
universality class and essentially mono-electronic methods such as
tight-binding, mean-field approximations or even density functional
methods fail to describe their electronic structure~\cite{dft}.  The
{\em active} electrons are energetically localized close to the Fermi
level and are thus responsible for the low energy properties of the
materials.  
It is thus convenient to partition the electrons of these strongly
correlated materials in a similar way as the usual core-valence
separation in atoms. On can define 2 classes, the {\em core}
electrons, essentially localized on the energetically deep,
closed-shell orbitals of the basic entities (atoms or molecules
according to the chemical nature of the crystal), and the previously
cited {\em active } electrons supported by localized, partially-filled
orbitals, presenting charge and/or spin fluctuations in the ground and
low energy-excited states.

While in finite, small systems, the treatment at the ab-initio level
of strong electron-correlation effects can be achieved in a
satisfactory manner, in infinite systems this is still an essentially
unsolved question.  The difficulty is even larger when excited states
are involved or thermodynamic properties seeked.  In order to by-pass
this problem and to understand the microscopic mechanisms underlying
the spectacular properties of these materials, physicists have
developed the usage of model Hamiltonians on the same bases than the
semi-empirical models used by quantum chemists in the $50$'s and
$60$'s~\cite{SEH} ---~Zero Differential Overlap (ZDO)
approximation~\cite{zdo} and reduction to the dominant
interactions~--- but with somewhat simpler expressions. Among the most
famous models one can cite the spin models such as the
Heisenberg-Dirac-Van Vleck models~\cite{heis1,heis2} for magnetic
half-filled systems, the $t-J$ models~\cite{t-J} for doped magnetic
systems, the Hubbard~\cite{hub} and extended Hubbard models
---~simplified versions of the Pariser-Parr-Pople~\cite{SEH}
approximation~--- that can describe metal-insulator phase
transitions. A simple description of these models is given in the
appendix.

The description and understanding of the electronic structure, low
energy physics and the associated macroscopic properties of these
materials is therefore achieved within a three steps procedure.
\begin{itemize}
\item The first step is the determination of an appropriate
model. Such a model, aimed at describing the low energy physics of the
material, must define the {\em active} electrons, the {\em active}
orbitals and the form of the dominant interactions between them.
\item The second step consists in the evaluation of the amplitude of
these interactions.  This evaluation can, in some well
conditioned cases, be extracted from experimental data, however in
most cases there is no unique and unambiguous way to determine such
parameters from spectroscopy or thermodynamic properties
measurements. 
\item The last step is devoted to the determination of the collective
properties of an assembly of electrons on a lattice, described by the
previously-defined, effective, local interactions .
\end{itemize}

We will see in this paper how it is possible to construct, step by
step, a coherent procedure for the determination of trustworthy
models describing the low energy physics of strongly correlated
materials. We will also see how quantum chemistry can play a major
role, both in the determination and validation of a model and in
reliable evaluations of the effective interactions amplitudes. The last
step is usually devoted to physicists since it requires the usage of
specific analytical methods such as renormalisation theories
bosonisation~\cite{boson}, Bethe
ansatz~\cite{bethe}, etc as well as specific numerical methods such as
Density Matrix Renormalisation Group (DMRG) methods~\cite{dmrg}, exact
diagonalisations, Monte-Carlo, etc. It aims at the
determination of the macroscopic order from the microscopic effective
model. It is however fascinating to follow the whole process and
understand the intrinsic link between the local chemical structure of
the materials and their macroscopic properties.

\section*{ \underline{\Large 2. Requirements for a {\em good}} \\ [0.7eM]
 \underline{\Large model}}
\label{sec:modreq}
Before any discussion on how to built an accurate and reliable model
for strongly correlated materials one should examine what means {\em
good}, {\em accurate} and {\em reliable}, that is what do we expect
from such a model. The model is supposed to fulfill several properties
that can be ordered into three sets~: the formal properties, the
physics we would like to reproduce and the transferability of the
interactions.

\subsection*{\underline{\large 2-A. Formal properties}}

These requirements are quite obvious and have already been
mentioned. They can be summarized as
\begin{itemize}
\item Hermiticity of the model Hamiltonian.
\item Small number of electrons and basis set orbitals per unit cell. 
\item Small number of basic interactions. 
\end{itemize}

\subsection*{\underline{\large 2-B. Modeling the low energy physics}}

The final aim of the model is to be able to produce a reliable
representation of the low energy physics and the properties of the
material, that is to be able to described and interpret not only phase
transitions but also experimental results such as low energy
spectroscopy, low energy photo-emission or photo-absorption, NMR,
X-ray or neutron diffraction, magnetic susceptibility, conductivity,
etc. We therefore need not only an accurate reproduction of the first
excitation energies but also accurate reproduction of the associated
wave-functions. In particular it is crucial to well reproduce the
charge and spin arrangements as well as the relative weight of the
major configurations in the ground and low energy excited states.
This means for instance, to reproduce accurately the ratio between the
coefficients of the Valence-Bond (VB) neutral and ionic
configurations, of the different spin arrangements within the neutral
configurations, etc. Such requirements on the wave
function suppose that \begin{itemize}
\item the low energy states we would like to  reproduce (let us
call them the target states $|\Psi_i\rangle,\;i=1\cdots m_t$) are
clearly defined (for instance states essentially based on neutral
Valence Bond configurations) and based on a small set of common active
orbitals,
\item the projection of the exact target states wave functions onto
the vector space supporting the model Hamiltonian (called the active
space from now on, $P$ being its projector operator) should be large
(one should notice that there is no requirement for the active
space to be complete)
\item the exact target states excitation energies ($\varepsilon_i,
\;i=1\cdots m_t$) should be well reproduced by the model hamiltonian
eigenvalues ($\epsilon_i$) or a convenient subset of them,
\item the model Hamiltonian eigenstates (noted
$|\xi_i\rangle,\;i=1\cdots m_a,\; m_a\ge m_t $), or a subset of them
should mimic the projection of the target states onto the active
space, that is in the ideal situation
$$ \forall i=1\cdots m_t, \quad |\xi_i\rangle = P|\Psi_i\rangle /
\sqrt{\langle \Psi_i|P|\Psi_i\rangle} $$
\end{itemize}

\subsection*{ \underline{\large 2-C. Locality of the interactions}}

As already mentioned, the model Hamiltonians are defined on a basis
set of local active orbitals and thus the microscopic interactions are
local, involve only a small number of electrons and a small number of
centers, usually one or two. Let us cite for instance, the
one-electron, two-centers hopping integral (as in the $t-J$ or
extended Hubbard models) or the two-electrons, two-centers
super-exchange integral (as in the $t-J$ or Heisenberg Hamiltonians).
One should however point out that the concepts of {\em center} or {\em
site} are not necessarily synonymous of {\em atom}, but rather the
expression of the basic building block participating to the low energy
physics of the system. For supra-conducting copper oxides for
instance, a center will be a copper atom, the oxygens and other atoms
being excluded from an explicit representation in the model.  For
molecular crystal a center will be the whole building molecule (such
as TMTSF~\footnote{tetramethyltetratiofulvalene},
TCNQ~\footnote{tetracyanoquinone}, etc) and its HOMO and/or LUMO will
be the localized active orbitals supporting the low energy physics of
the materials. The concept of locality can thus refer to fairly large
units from the molecular point of view but of size always inferior to
the unit cell and therefore localized from the crystal point of view.
The locality assumption has the immediate and important consequence
that the interactions do not depend of the number of explicit active
centers and that, provide a set of orbitals and associated energies, a
$n$-centers interaction will act in a finite system in the same way as
in an infinite one. This property is usually called
transferability. The hopping or the exchange integrals can, for
instance, be defined from the low excitation energies of a two centers
system. The exchange integral is the singlet triplet energy difference
in a two active electrons system and the hopping integral is half the
symmetric versus antisymmetric doublet energy difference, in a single
active particle system.

\begin{center}
\begin{center}
{\bf The exchange integral $ J_{ab} $} \\
 2 ${\bar e}$ - 2 orb. - 2 sites 
$${{\uparrow \atop \bullet }\atop a}{{{\longrightarrow \atop \longleftarrow }\atop
 \hfill }\atop \hfill }{{\downarrow \atop \bullet }\atop b}$$ 
${\bf J_{ab}}=E(sg)-E(tp)$ 

\vspace*{2eM}

{\bf Hopping integral $ t_{ab} $} \\
 1 ${\bar e}$ - 2 orb. - 2 sites 
$${{\uparrow \atop \bullet }\atop a}{{\longrightarrow \atop
 \hfill }\atop \hfill }{{\hfill\atop \bullet }\atop b}$$ 
${\bf t_{ab}}=\left(E(db_+)-E(db_-)\right)/2$ 
\end{center}
\end{center}

Despite their local definition in the model Hamiltonians scheme, it is
clear that these interactions are {\em effective} and one should
wonder whether they are truly local or whether they do involve in an
effective manner non local mechanisms. This is the question behind the
transferability problem. Its answer is crucial for the determination
of reliable models. Indeed, without the assurance of a true locality
of the interactions the whole model hamiltonian scheme fails.  The
derivation of the macroscopic properties of an infinite lattice is
based on the idea that this infinite system is the limit of increasing
large size finite systems. The implicit hypotheses behind it are that
the interactions involve a finite and small number of bodies and that
their nature and amplitude do not change when the system size
increases, i.e.  they are perfectly transferable.  The effective
inclusion of non local mechanism is incompatible with these hypotheses
since it would render the effective integrals dependent on the system
size.

One should however remember that the exact Hamiltonian can always be
expressed in an atomic basis set where all interactions are local and
transferable. The non locality of a model Hamiltonian can therefore
always be fixed by an appropriate modification of the model, usually
consisting in the addition of longer range interactions, interactions
involving a larger number of centers or an enlarged definition of the
active space. The transferability property is therefore an useful tool
for the verification of the model pertinence. A model which is not
transferable usually neglects the explicit treatment of an important
physical effect and the transferability is restored by the inclusion
of the omited mechanism within the model Hamiltonian.  The whole
process is however limited by the complexity of the final model, the
number of basic interactions and the number of bodies involved in it.

\section*{\underline{\Large 3. Locality of effective}\\  
[0.7eM] 
\underline{\Large interactions}}
\label{sec:loc}

In this section we will investigate the locality problem for the major
types of effective interactions found in model Hamiltonians, namely
hopping, exchange and repulsion. In this analysis we will suppose that
the orbital space has been partitioned into the
\begin{description}
\item[core orbitals], these orbitals are essentially doubly-occupied
in the target states and support the inactive electrons, explicit
reference to them is usually omitted in the model Hamiltonian,
\item[active orbitals], they support the active electrons responsible
for the low energy physics, these orbitals are supporting the model
Hamiltonian and have fluctuating occupation and/or spins in the target states,
\item[virtual orbitals], these orbitals are essentially empty in the
target states and omitted in the model.
\end{description}
The above orbital basis set will be supposed to be local, orthogonal
and the orbitals composition (as can be given by their expansion on
the atomic orbitals basis set) and energies already defined. We will
come back later on the problem of these orbital and orbital energies
definitions and analyze the non-locality contained in it.  This
partition of the all electrons basis set is crucial and will be
refered at all over this paper.

The effective integrals can be decomposed (using the above partition
of the orbitals basis set) into a {\em zeroth-order} part
corresponding to the exact Hamiltonian integrals within the active
space, and a higher order part that takes effectively into
account the effect of the configurations out of the active space. It
is in this higher order part that non local mechanisms may be
introduced through long-range dynamical polarisation or correlation
effects.

We will start by the analysis of the effective bi-electronic integrals
since their case is somewhat simpler than for the mono-electronic
ones. 

\subsection*{ \underline{\large 3-A. The effective exchange integrals}}

The effective exchange integrals appear in magnetic models such as the
Heisenberg or $t-J$ Hamiltonians. These models are suited for the
description of very strongly correlated systems where the on-site
bi-electronic repulsion is so high that the explicit reference to
configurations where an active orbital is doubly occupied can be
excluded from the model. The essential role of the effective exchange
integrals is therefore to take into account the effects of these
double occupancies on the interactions between neutral Valence-Bond 
configurations, that is to position properly the local
singlet to triplet excitation energy  between two nearby centers.

\begin{figure}[h]
\resizebox{8cm}{3cm}{\includegraphics{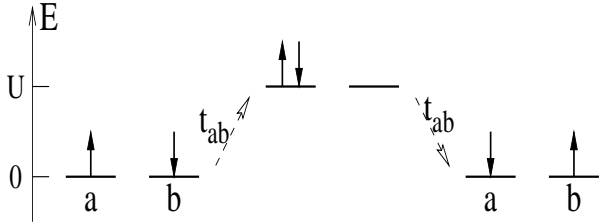} }
\vspace*{0.5eM}
\caption{Anderson super-exchange mechanism}
\label{fig:and}
\end{figure}

Within the complete configurations space defined from the active
orbitals (CAS), Anderson~\cite{and} has clearly detailed this
mechanism and provided a perturbative expression of the effective
super-exchange integral $J_{ab}$ (see figure~\ref{fig:and}) \\

\begin{eqnarray*} \label{eq:and}
J_{ab} &=& -4 {\langle a \bar b| H | a \bar a \rangle^2 \over 
E\left(|a \bar a\rangle\right) -  E\left(|a \bar b \rangle\right) }\\
&=& - 4 {t_{ab}^2 \over U}
\end{eqnarray*}
where $t_{ab}$ is the hopping between the magnetic orbitals $a$ and
$b$ and $U$ is the Coulomb repulsion integral when a magnetic orbital
is doubly-occupied.

\begin{figure}[h]
\hspace*{-1.0cm}\resizebox{10cm}{4.3cm}{\includegraphics{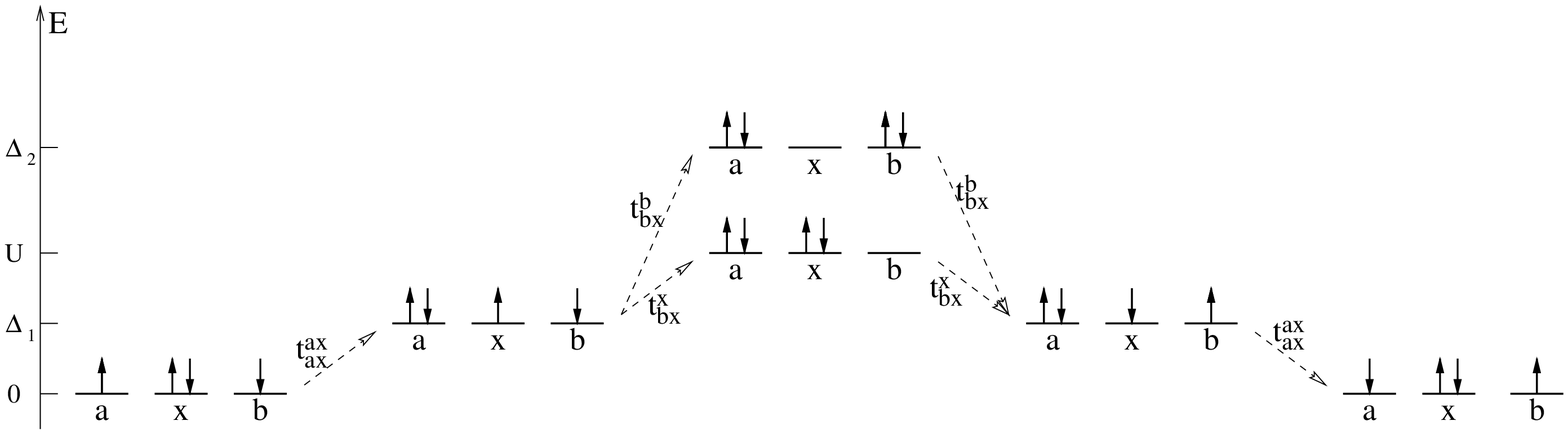} }
\vspace*{0.5eM}
\caption{Through-bridge super-exchange mechanism. In this example we
supposed that the process goes through the occupied orbitals of the
bridge, represented by x. It may also go through the unoccupied
orbitals or both.}
\label{fig:pontJ}
\end{figure}

Things are a little more complicated when the exchange mechanism is
not direct but through a chemical bridge. This is for instance the
case in most transition metal oxides where the super-exchange
interaction between two $d$ orbitals of nearby metal atoms is mediated
through the oxygen atom(s). In this case the leading perturbative term
comes at the fourth order and two different mechanisms add (see
figure~\ref{fig:pontJ}).

\begin{figure}[h]
\resizebox{8.0cm}{3cm}{\includegraphics{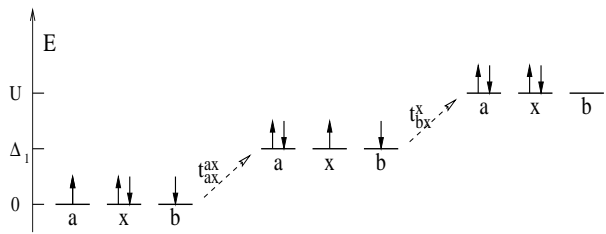} }
\vspace*{0.5eM}
\caption{Through-bridge hopping mechanism. In this example we supposed
that the process goes through the occupied orbitals of the bridge,
represented by x. It may also go through the unoccupied orbitals or
both. }
\label{fig:pontt}
\end{figure}

\begin{eqnarray*}
 J_{ab} &=& -4 {\left(t_{ax}^{ax}\right)^2  \left(t_{bx}^x\right)^2
  \over \left(\Delta_1\right)^2  U} 
-8{\left(t_{ax}^{ax}\right)^2 \left(t_{bx}^b\right)^2 \over 
\left(\Delta_1\right)^2  \Delta_2}
\end{eqnarray*}
where 
\begin{eqnarray*}
t_{bx}^x &=& \langle x \bar x |H| x \bar b  \rangle \; 
\text{ is the hopping between orbitals } \\
&&  b \text{ and } x 
\text{ with a spectator electron on } x \\
t_{bx}^b &=& \langle b \bar x |H| b \bar b  \rangle  \; 
\text{ is the hopping between orbitals } \\
&&  b \text{ and } x 
\text{ with a spectator electron on } b \\
t_{ax}^{ax} &=& \langle a x \bar x |H| a x \bar a \rangle   \; 
\text{ is the hopping between orbitals } \\
&&  a \text{ and } x 
\text{ with two spectators electron on } a \text{ and } x \\
\Delta_1 &=& E\left(|ax^2\bar b\rangle\right) - E\left(|a^2x\bar
b\rangle\right) \\
U&=& E\left(|ax^2\bar b\rangle\right) - E\left(|a^2x^2\rangle\right) \\
\Delta_2 &=& E\left(|ax^2\bar b\rangle\right) - E\left(|a^2 b^2\rangle\right)
\end{eqnarray*}
  It should be noted that the second path goes through a configuration
where the doubly-occupied orbital of the bridge has been totally
emptied (or totally doubly-occupied if the process goes through the
unoccupied orbitals of the bridge). Such a configuration is usually
very energetic and its contribution negligible in front of the first
path. In that case the through-bridge super exchange integral can
again be written in an Anderson's form $J_{ab} = 4 t_{ab}^2/U$
provided that an effective, through-bridge, hopping integral $ t_{ab}$
between the two magnetic orbitals is defined as (see
figure~\ref{fig:pontt}) 
$$ t_{ab} = {t_{ax}^{ax} t_{bx}^x \over \Delta_1}$$ 

\begin{center}
\begin{figure}[h]
\resizebox{5cm}{2.5cm}{\includegraphics{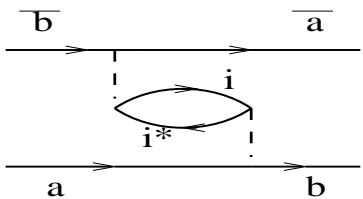} } 
\vspace*{0.5eM} 
\caption{Leading non local contributions to the exchange effective
integral. $i$ and $i^*$ are respectively  occupied and vacant orbitals 
on a distant site.}
\label{fig:expol}
\end{figure}
\end{center}

It is clear that all these processes included in the effective
exchange integral are totally local. Dynamical polarisation or
correlation processes that may act on top of the them (for a review of
all diagrams contributing to the unbridged super-exchange mechanisms
at the second order of perturbation see ref.~\cite{deLoth}) remain
local as long as they involve only orbitals located on the magnetic or
neighboring atoms, including the possible bridge.  The only
possibility to include non-local contributions is therefore the
dynamical polarisation and correlation processes where the excitations
take place on a center far from the magnetic orbitals. The leading
contributions of such terms (second order of perturbation) correspond
to the super-exchange plus polarisation diagrams in the DeLoth
denomination~\cite{deLoth} (see figure~\ref{fig:expol})

The corresponding perturbative  contribution is  
$$- \frac{\langle i \bar b|{1\over r_{12}}|i^*\bar a \rangle \langle
i^* a|{1\over r_{12}}|ib \rangle}{\Delta E}$$ which varies as a
function of the distance $R$ between the magnetic distribution $ab$
and the distant local excitation $ii^*$ as a dipole-dipole
interaction, that is as $1/\left(R^3\right)^2 = 1/R^6$.

One can therefore reasonably consider that these non-local
contributions are negligible and that the exchange effective integral
is really local. This locality property has been numerically verified
by Illas and coworkers~\cite{XJloc} from ab-initio calculations on a
variety of compounds such as the $\rm KMF_3$ and $\rm K_2MF_4$ ($\rm
M=Cu,Ni$) family or hight $T_c$ perovskites.

\subsection*{ \underline{\large 3-B. The effective coulomb repulsion integrals}}

The effective coulomb repulsion terms act on the diagonal of the
effective Hamiltonian. They correspond to screened bi-electronic
repulsions. This is through this screening that non local effects may
occur.

The screening that may come from the active electrons is treated
explicitly in the model Hamiltonian, for instance through hybridisation
and delocalisation processes, and therefore do not contribute to the
effective repulsion integral. It remains the dynamical polarisation
and correlation effects.

The main non-local contributions come from dynamical correlation
effects through double-excitations from two core orbitals towards two
virtual ones. However, these excitations yield equal contributions
(at the second order of perturbation) to all VB configurations in the
active space and give only a shift in the definition of the model
Hamiltonian zero of energy. Thus they are non pertinent and can 
be omitted.

\begin{center}
\begin{figure}[h]
\resizebox{5cm}{2.5cm}{\includegraphics{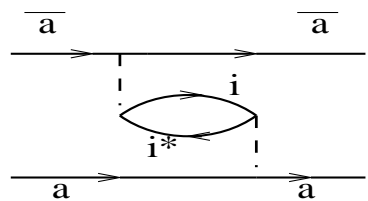} }
\vspace*{0.5eM}
\caption{Leading non local contribution to the on-site effective
repulsion integral between two electrons in the same orbital. $i$ and
$i^*$ are respectively occupied and vacant orbitals of the distant
excitation}
\label{fig:Upol}
\end{figure}
\end{center}

The remaining leading non-local contributions come from distant
single-excitations. For the sake of simplicity, we will suppose that
the core ($i$) and virtual ($i^*$) orbitals are eigenvectors of the
core Fock operator (as defined from the antisymmetrized product of all
doubly-occupied core orbitals). The leading term comes therefore from
the coupling between the $(i\,i^*)$ distant distribution and the local
$(a\,\bar a)$ distribution (if the two electrons are on the same
orbital), $(a\,\bar b)$ or $(a\,b)$ distributions (if the two
electrons are on different orbitals). See figure~\ref{fig:Upol}).

In the case where the effective repulsion integral acts between two
electrons in the same orbital, the leading non-local contributions are at the second order contribution 
$$- \frac{\langle i \bar a|{1\over r_{12}}|i^*\bar a \rangle \langle
i^* a|{1\over r_{12}}|ia \rangle}{\Delta E}$$ 
This terms vary as a
function of the distance $R$ between the $a \bar a$ distribution and
the distant excitations $ii^*$ as a charge to dipole interaction,
that is as $1/\left(R^2\right)^2 = 1/R^4$. 

When the repulsion integral acts between  two electrons in different
orbitals, this dominant non-local contributions vary as a
dipole-dipole interactions, that is as $1/\left(R^3\right)^2 = 1/R^6$.

The non-local contributions to the repulsion integrals vary
therefore either as $1/R^4$ or as $1/R^6$, that is 
fast enough  to consider that these effective integrals are
essentially local. 

\subsection*{ \underline{\large 3-C. The effective hopping integrals}}

The case of hopping integrals or other mono-electronic integrals (such
as orbital energies for instance) is somewhat more complicated. Indeed
in their zeroth-order description (projected exact Hamiltonian in the
active space) these integrals are coupled to the Fock operator, that
is both to the nuclear and the electronic charge repartition on the
entire system. It is clear that, through this coupling, the
mono-electronic integrals are non local. However, once provided the
definition of the orbitals (core, active and virtual), their
composition in terms of the atomic orbitals and their energies, the
other contributions to the hopping integrals can be analyzed as was
done for the super-exchange or repulsion integrals. Let us decompose
the problem into three terms, the first term is the coupling with the
core orbitals (and the corresponding charge density of the entire
system) through the Fock operator, the second term is contributions
from the active orbitals, the third term is the excitations out of the
active space that are included in an effective manner.
 
The first term is totally defined by the Fock operator. The problem
intrinsically depends upon the average charge repartition on the
entire system and is solved by the definition of the orbitals
themselves through a self-consistent, mean-field approximation. This
is the major non local contribution and its consequences is that the
infinite system is never really treated as the limit of {\em true}
large but finite system ---~since the edge effects should not be taken
into account in the orbitals and orbital energies definitions. A
proper definition of the orbitals therefore suppose that the finite
systems are in fact embedded in a mean field approximation of the rest
of the crystal.

The second term reminds us that the hopping integrals are dependent on
the specific configurations they act upon.  For instance the transfer
of an electron on a bond depend on the number and the nature of the
spectators (active) electrons on it.

\begin{center}
\begin{figure}[h]
\resizebox{5.5cm}{1.5cm}{\includegraphics{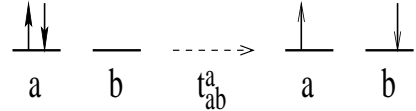} }
\vspace*{0.5eM}
\caption{$t^a_{ab}$ hopping with a spectator electron in $a$.}
\label{fig:ta}
\end{figure}
\end{center}

\begin{eqnarray*}
 t^{a}_{ab} &=& \langle a  \bar a |H| a  \bar b \rangle \\&=& 
\langle \bar a |H|\bar b \rangle + 
\langle a \bar a | 1/r_{12}| a  \bar b \rangle 
\end{eqnarray*}

\begin{center}
\begin{figure}[h]
\resizebox{5.5cm}{1.5cm}{\includegraphics{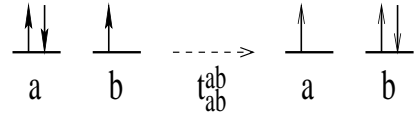} }
\vspace*{0.5eM}
\caption{$t^{ab}_{ab}$ hopping with two spectator electrons in $a$ and
$b$.}
\label{fig:tab}
\end{figure}
\end{center}

\begin{eqnarray*}
 t^{ab}_{ab} &=& \langle a b \bar a |H| a b \bar b \rangle \\&=& 
\langle \bar a |H|\bar b \rangle + \langle a \bar a | 1/r_{12}| a  \bar b \rangle + \langle b \bar a | 1/r_{12}| b  \bar b \rangle 
\end{eqnarray*}

The hopping integrals therefore depend upon the distant active space
charge distribution of the configurations they act on. The
corresponding contributions vary as $\langle x a |1/r_{12}-
xb\rangle$, that is as $1/R^2$ if $R$ is the distance between the
$(ab)$ and the distant $(xx)$ distributions. This dependence should be
explicitly include in the model hamiltonian definition. However,
despite its rather slow decrease with $R$, it is usually ignored, 
 except may be for the dependence to the occupation of
the orbitals localized on the same sites as $a$ and $b$.

\begin{center}
\begin{figure}[h]
\resizebox{5cm}{2.5cm}{\includegraphics{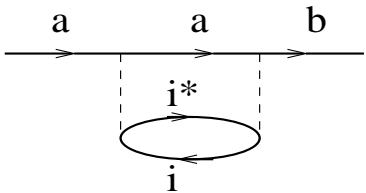} }
\vspace*{0.5eM}
\caption{Leading non local dynamical contribution to the effective
hopping $t_{ab}$.}
\label{fig:tpol}
\end{figure}
\end{center}

The third term contributions are included in an effective manner.
Their leading non-local processes come from distant single-excitations
acting on the transfer, as shown in figure~\ref{fig:tpol}.

The Fock operator being defined as previously, the  perturbative expression of the dominant is 
\begin{eqnarray*}
&-& {\langle a i |1/r_{12}| a i^* \rangle \langle a i^*|1/r_{12}| b i\rangle \over \Delta E_a} \\
&-& {\langle a i |1/r_{12}| b i^* \rangle \langle b i^*|1/r_{12}| b i\rangle \over \Delta E_b}
\end{eqnarray*}
which varies as $1/R^2 \times 1/R^3 = 1/R^5$, that is fast enough to
be considered as a non-significant source of effective non-local
processes.

\vspace*{2eM}

As a summary the major types of interactions encountered in model
Hamiltonians are essentially local in nature, provided that the
coupling to the average charge repartition of the entire crystal
(which is the major non-local mechanisms) is treated through a proper
definition of the orbitals and orbital energies. It is therefore
pertinent to evaluate the interactions amplitudes and check a model
validity by ab-initio calculations on  properly embedded
fragments. Such a technique presents a lot of  advantages.
\begin{itemize}
\item All short range effects can be explicit and accurate treated by
the inclusion of the active sites local environment in the computed
fragment.
\item The dynamical processes that strongly affect the values of the
effective interactions can be accurately evaluated. Indeed, it is
possible to use large atomic basis set and up-to-date quantum chemical
methods for the fragment spectroscopy.
\item One has access not only to the local excitation energies but
also to the associated wave functions. In particular the knowledge of
the relative contributions of the major active space VB configurations
is determinant for the good reproduction of the material properties by the
model.
\end{itemize}
The question of the embedding is therefore crucial, as it determines
the composition of the orbitals (core, active and virtuals) as well as
their energies.

\section*{ \underline{\Large 4. How to built a {\em good} bath}}
\label{sec:bain}

According to the previous analysis a {\em good} embedding should be
such that the fragment orbitals and their energies are similar to what
they would be in the crystal. Reaching this goal necessitates that the
effects of the rest of the crystal on the fragment are properly
reproduced. These effects may be partitioned into long range effects
and short range effects.

The previous analysis showed us that the major long range effects are
mono-electronic, comes through the Fock operator definition and more
specifically through the coupling to the entire system average charge
repartition. Considering that these strongly correlated systems are
most of the time ionic, this term is large and corresponds to the
Madelung potential. It can be easily reproduced by a set of positive
and negative charges respectively positioned at the location of the
cations and anions of the rest of the crystal. The volume involved
should be large enough and centered on the fragment.  Border charges
can be adapted according to the Evjen procedure in order
to increase the potential convergence.

The short range effects of the rest of the crystal on the fragment are
essentially exclusion effects. They come from the orthogonality
requirement of the fragment orbitals to the orbitals of the rest of
the crystal. This effect is crucial since in the vacuum the orbitals
of the fragment ---~which is often highly negatively charged~--- have
a tendency to extend far out of the fragment volume, in region of
space where the other electrons of the crystal should be located. It is
therefore necessary to forbid these domains of ${\cal R}^3$ by a
repulsive potential mimicking the electronic volume of the atoms of
the rest of the crystal. This aim can be reached by the usage of total
ions pseudo-potentials~\cite{tips} at the location of the first shells
neighboring atoms of the fragment.

Since all these embedding effects are mono-electronic they can be
easily checked. Indeed, they are included in a simple Hartree-Fock
(HF) calculation thus a good embedding should reproduce the HF density
of states of the entire crystal, projected onto the fragment atomic
orbitals.

\subsection*{ \underline{\large 4-A. Example~: the $\alpha^\prime N\!aV_2O_5$ compound}}

\begin{figure}[h]
\resizebox{8cm}{5cm}{\includegraphics{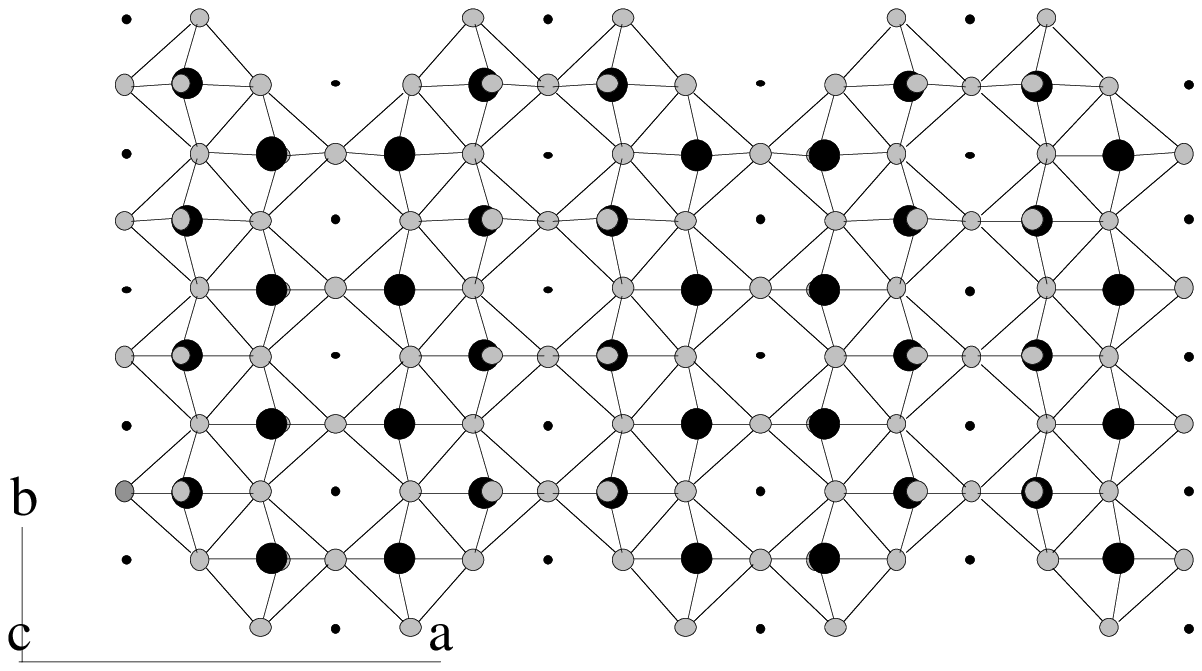} }
\resizebox{8cm}{5cm}{\includegraphics{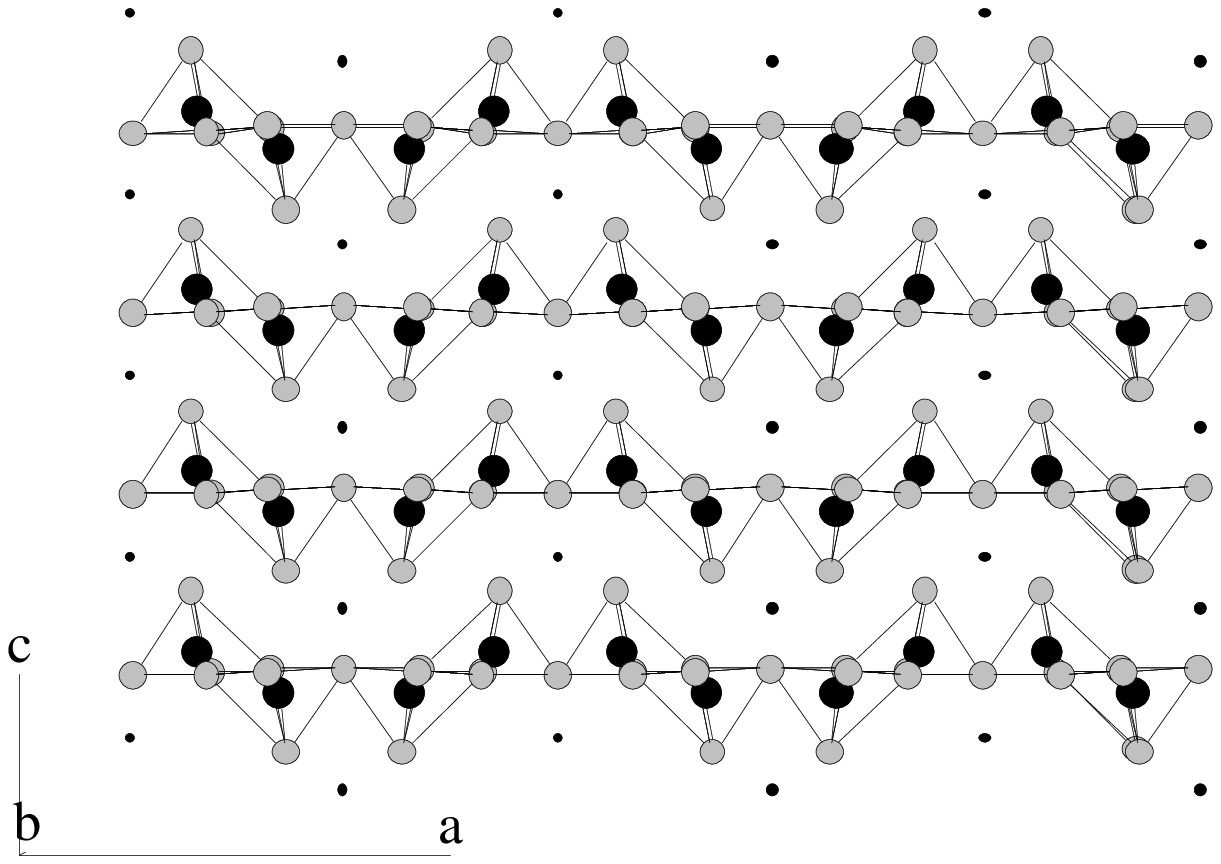} }
\vspace*{0.5eM}
\caption{Schematic structure of $\alpha^{'}N\!aV_2O_5$, along the
a-b and  the a-c planes. The oxygen atoms are  denoted by
open circles, the vanadium atoms by filled circles and the sodium
atoms by dots.}
\label{fig:vana-struct}
\end{figure}

As an example of the importance of the embedding on the orbitals
definitions and orbitals energies~\cite{env} we will discuss the case of the
$\alpha^\prime N\!aV_2O_5$ compound. $\alpha^\prime N\!aV_2O_5$ is a
ionic layered compound built from $VO_5$ square-pyramids arranged in
$(a,b)$ planes (see figure~\ref{fig:vana-struct}) and doped by sodium
ions located between the layers.

\begin{figure}[h]
\hspace{-0.3cm}\resizebox{9cm}{5cm}{\includegraphics{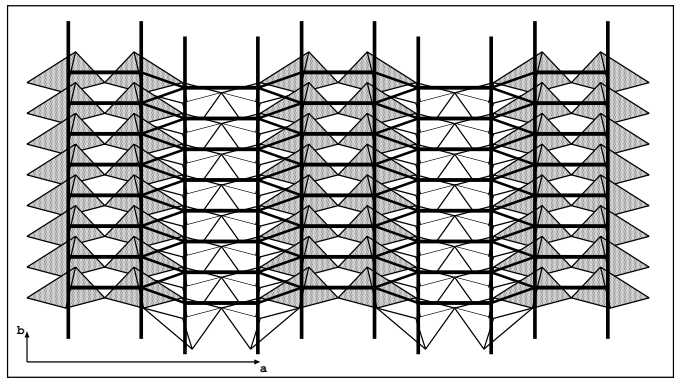} }
\vspace*{0.5eM}
\caption{ Schematic structure of the coupled magnetic ladders in the
(a,b) plane.}
\label{fig:vana-ech}
\end{figure}

The compound is strongly ionic and the formal charge transfer yields a
formula of $N\!a^+\left(V_2\right)^{9+}\left(O^{2-}\right)_5$. The
vanadium atoms support therefore one unpaired electron for two
magnetic sites. It has been shown that the different layers are very
weakly coupled and that the electronic structure can be considered as
two dimensional~\cite{Galy2}. In fact most authors see the compound as
coupled, quarter-filled (one active electron for two active orbitals),
magnetic ladders (see figure~\ref{fig:vana-ech}).

Let us concentrate, for instance, on the interactions of the magnetic
electrons on a rung of the ladders (for an ab initio evaluation of the
different interactions in the high temperature phase see
ref.~\cite{vana1}, in the low temperature phase see ref.~\cite{vana3}
and for an ab-initio evaluation of the charge ordering
ref.~\cite{vana2}). They are of two types, hopping and
super-exchange. They can be evaluated by the spectroscopy of an
embedded fragment composed of two $VO_5$ pyramids sharing a corner
oxygen (see figures~\ref{fig:vana-rungab} and \ref{fig:vana-rungac}).
\begin{center}
\begin{figure}[h]
\resizebox{5cm}{2.5cm}{\includegraphics{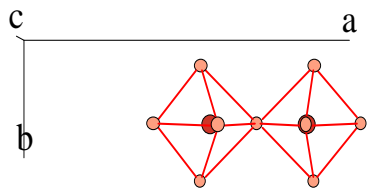} }
\vspace*{0.5eM}
\caption{Fragment of the crystal used for the cluster calculations of
the interactions along a rung seen in the $(a,b)$ plane.}
\label{fig:vana-rungab}
\end{figure}

\begin{figure}[h]
\resizebox{5cm}{2.5cm}{\includegraphics{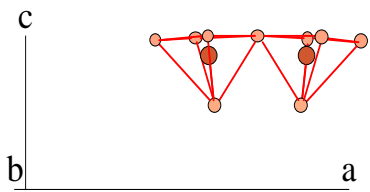} }
\vspace*{0.5eM}
\caption{Fragment of the crystal used for the cluster calculations of
the interactions along a rung seen in the $(a,c)$ plane. }
\label{fig:vana-rungac}
\end{figure}
\end{center}

\begin{figure}[h]
\begin{minipage}{4.cm}
\centerline{\resizebox{3cm}{6cm}{\includegraphics{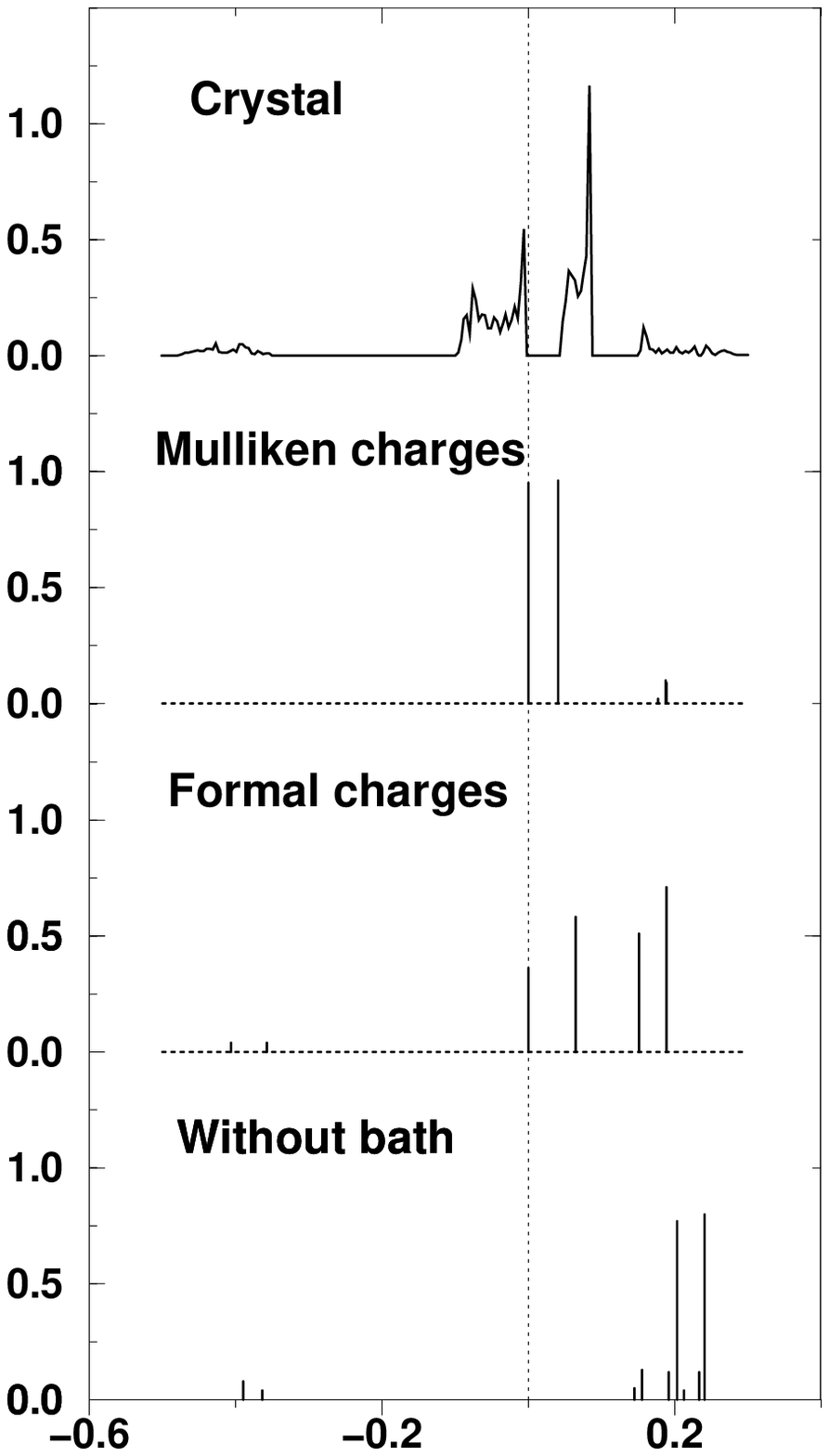} }}
\vspace*{0.5eM}
\caption{Density of states projected on the magnetic $d_{xy}$ orbitals
of the rung vanadium atoms~\cite{env}.}
\label{fig:vana-dosd}
\end{minipage}   \hfill
\begin{minipage}{4.cm}
\centerline{\resizebox{3cm}{6cm}{\includegraphics{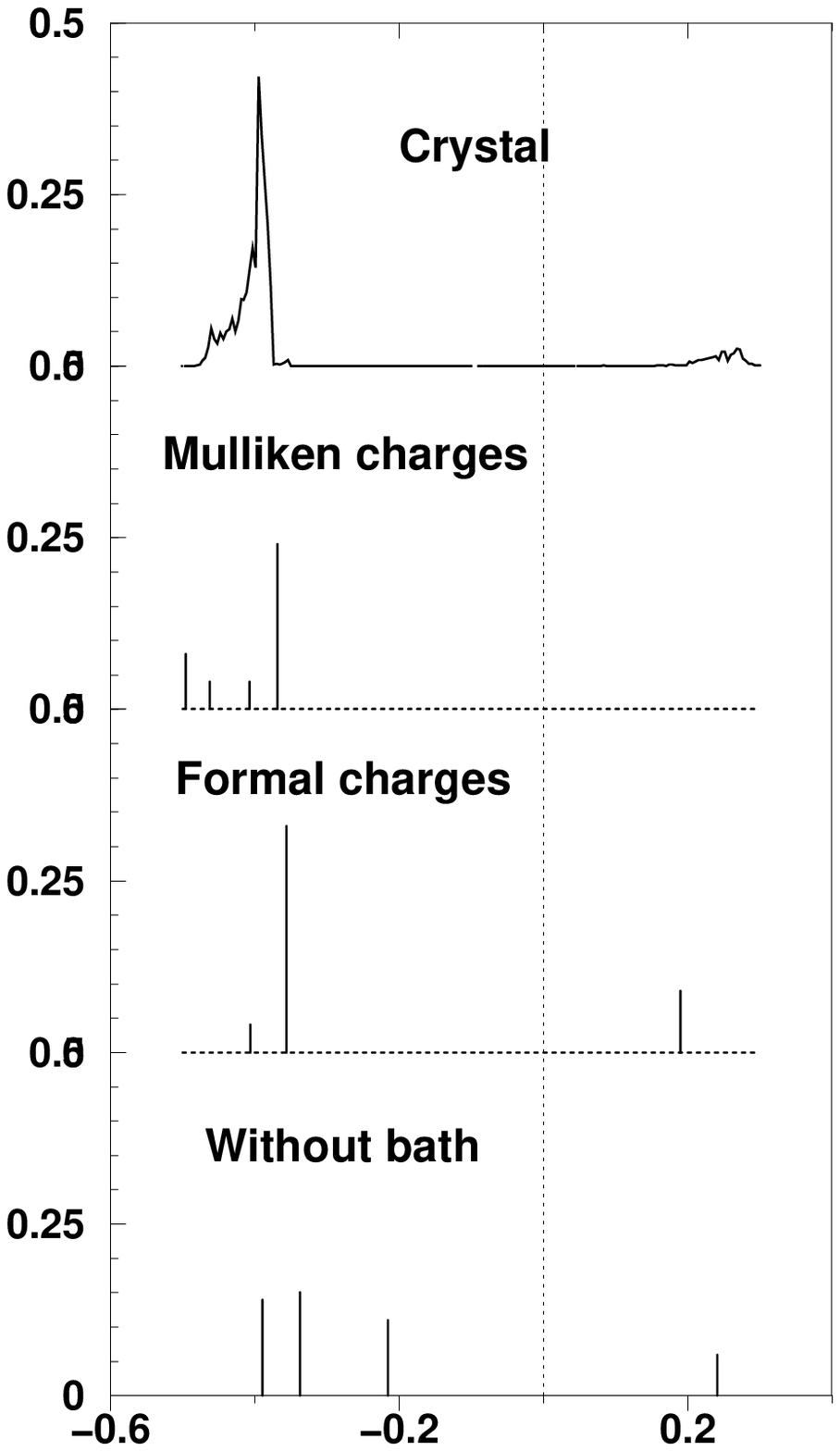} }}
\vspace*{0.5eM}
\caption{Density of states projected on the $2p_y$ orbital of the 
rung bridging oxygen atom~\cite{env}.}
\label{fig:vana-dosp}
\end{minipage}  
\end{figure}

We can therefore check  the importance of the embedding on the
orbitals definitions and orbitals energies by the comparison of the
density of states (DOS) of the entire crystal (as computed from
periodic HF techniques~\cite{Crystal}), the density of states of the
isolate fragment and the density of states of the fragment embedded in
different baths.  For sake of clarity we will project the DOS on the
most important orbitals for the effective hopping and exchange
interactions, namely the magnetic orbitals of the vanadium atoms
($3d_{xy}$ if $x,y,z$ refer respectively to the crystallographic
orthogonal $a,b,c$ directions) and the $2p_y$ orbital of the bridging
oxygen that mediate the interactions. 
Figure~\ref{fig:vana-dosd} and figure~\ref{fig:vana-dosp} report
respectively the different projected DOS on the magnetic and bridging
orbitals. One can see immediately that the isolated fragment DOS is
very different from the infinite crystal one. In fact, while the
highest occupied crystalline bands are built from the $d_{xy}$ atomic
orbitals of the vanadium atoms, in accordance with the magnetic
character of this orbital, in the isolated fragment the molecular
orbital based on the vanadium $3d_{xy}$ is far above the Fermi level,
the unpaired Highest Occupied Molecular orbital being a delocalized
orbital on the $3p$ orbitals of the vanadium and the $2p$ orbitals of
different oxygen atoms.  The embedding built from atomic formal
charges ($V^{4.5+}$, $O^{2-}$ and $N\!a^+$) is properly reproducing
the DOS projected on the oxygen $2p_y$ bridging orbital, however it is
unable to reproduce the crystal DOS projected on the magnetic
orbitals. An orbital analysis of the different embedded
fragments and crystal calculations shows that the apical oxygen is not
linked to the vanadium by an ionic bond but that a strong, dative,
triple bond is taking place between these two atoms. The consequence
is a strong deviation of the Madelung potential from the one defined
by formal charges. A redefinition of the bath using approximate
self-consistent M\"ulliken charges on the vanadyle, that is $V^{3+}$
and $O^{0.5-}$ finally yield a proper reproduction of the DOS
projected on all the orbitals of the fragment, insuring a reliable
taking into account of the non-local crystal effects.

\vspace*{1eM} Let us now take a look at the importance of these
embedding problems on the effective integrals amplitudes, namely the
hopping $t_\bot$ integral of a magnetic electron between the two
vanadium atoms of the rung, and the super-exchange $J_\bot$ integral
of two magnetic electrons on the rung. As seen previously, these
integrals can be defined from the fragment spectroscopy as half the
symmetric versus antisymmetric doublet excitation energy for the
hopping ($t_\bot = 1/2 \left[ E(D_g) - E(D_u) \right]$), and as the
singlet versus triplet excitation energy for the exchange ($J_\bot =
E(Sg) - E(Tp)$). We shall not describe here the computational
technique used for these spectroscopic calculations since the next
section will be devoted to this problem. Table~1
integrals both for the formal charges and the Madelung charges
embeddings.  One can see immediately that a small change in the dipole
moment of the vanadyle fragment ($V^{4.5+}O^{2-}$ in the formal
charges bath, $V^{3+}O^{0.5-}$ in the so-called M\"ulliken charges
bath) can considerably modify the effective integrals amplitudes, since
$J_\bot$ is multiplied by a $1.73$ factor between the two
calculations.

\begin{center}
\begin{tabular}{r|rr} \hline \hline
	Integral & Mulliken charges & Formal charges     \\
\hline
$J_\bot$ & -293.5 meV & -509.7 meV  \\
$t_\bot$ & -538.2 meV & -581.2 meV  \\ \hline \hline
\end{tabular}  \vspace*{1eM}
\end{center}
{\normalsize TABLE 1. Hopping and super-exchange effective integrals
of magnetic electrons between the two vanadium sites of the rung
according to the nature of the embedding~\cite{env}. }

\section*{ \underline{\Large 5. Fragment spectroscopy}}

The fragment embedding being properly defined, the question is now how
to computed accurate excitation energies and related 
wave-functions. The question is far from trivial since the total
energies of the computed states are usually of the order of $10^3eV$
while the seeked transition energies are of the order of a few tenth
up to a few tens of $meV$, that is from $5$ to $7$ orders of magnitude
smaller. Even the correlation part of the energies are of the order of
a few tens of $eV$ which is still between $3$ to $5$ orders of
magnitude larger than the seeked excitations energies. Reliable
results therefore necessitate procedures that calculate
identically the major part of the total energies. This can only be
achieved if a common set of unbiased orbitals is used for the
different states involved in the excitation, whether they do belong to
the same irreducible symmetry representation (space and spin) or they
do not. 

A zeroth-order description of these seeked target states can be
obtained using a CASCI~\footnote{Complete Active Space Configurations
interaction} calculation. The main question is the choice of the set
of orbitals supporting the CAS space. While it is obvious that the
active orbitals (as defined in sections~3 
and~1) should belong to this set, one may wonder whether
it should be restricted to them. The question is specially crucial in
the case of bridged systems where the interactions are mediated by a
ligand and the question of including in the active space the mediating
orbitals of the bridge (denoted as $x$) should be asked.  The answer
depends on the weight ---~in the target states~--- of the
configurations involving a partial occupation of $x$~; that is
different from $2$ if $x$ is a core orbital and different from $0$ if
$x$ is a virtual one.  If large, the mediating orbitals $x$ should be
included into the set supporting the CAS.
In such cases, one may wonder whether the definition of the active
orbitals supporting the model shouldn't be enlarged
accordingly. Independently of the definition of the CAS for the
ab-initio calculations, the answer to this question and accordingly
the validity of the model and related active orbital space
definitions, can be checked using a simple criterium~: the norm of
the projection onto the active space supporting the model
Hamiltonian, of the computed target states wave functions. A large
value ($\ge 0.7$) of this projection insures a correct definition of
the model while a small value is the sign that an important physical
effect have been omitted.

While the CAS is a good reference description, it is well known that a
CASCI calculation is far from yielding accurate excitation
energies. Indeed it is missing not only the dynamical correlation
effects, but also the space and spin repolarisation effects specific
to the different VB configurations (dynamical polarisation), which are
known to be crucial both for a proper treatment of charge
delocalisation~\cite{poldyn} and for correct relative weights of the
different VB configurations (in particular the ionic ones) in the
active space~\cite{pd}.
This last point is of great interest since it discards all
computational methods which do not allow a revision of the relative
weights of the different VB configurations within the CAS, in
perspective of the dynamical polarisation effects. This is the case
for instance of the perturbative methods after a CAS zeroth order. 
They necessitate large CAS including the major part of the
repolarisation effects in the zeroth order wave function in order to
give accurate results~\cite{CASPT2}.

As a matter of illustration table~2 reports the evaluation of the
effective exchange and hopping integrals on the rung of the
$\alpha^\prime N\!aV_2O_5$ compound computed both at the CASCI (the
CAS being supported by the magnetic orbitals and the mediating orbital
of the bridging oxygen) and the CASCI + all single excitations on all
CAS determinants (zeroth-order + dynamical polarisation) levels.  One
notices immediately that the effective integrals are dramatically
modified by the inclusion of the dynamical polarisation
effects. Indeed their inclusion increases the super-exchange integral
by a factor $\simeq 5.3$ and the effective hopping by a factor of
$\simeq 1.3$. One may however wonder why this repolarisation effect is
so much stronger on the exchange integral than on the hopping. At the
simplest level of description, the hopping involves a charge
fluctuation between the two magnetic sites, thus strong changes in the
electrostatic potential. An accordingly strong dynamical polarisation
response can be expected. On the contrary the exchange integrals
involves only spin field modifications which is usually expected to
yields much smaller repolarisation effects. This apparent
contradiction can be levelled if one remembers the through-bridge
super-exchange mechanism shown in figure~\ref{fig:pontJ}. While the
through-bridge hopping mechanism involves only paths going through
configurations with a $1 \bar e$ occupation change (relative to the
references) of the bridging orbital $x$, the exchange mechanism
involves not only similar paths, but also paths where the bridging
orbital occupation is modified by $2 \bar e$. This is these paths that
induce extremely strong specific repolarisation effects. Indeed, in
the present case this last path carries half of the effective exchange
amplitude explaining the observed much stronger repolarisation effects
than on the hopping integral.

\vspace*{1eM}
\begin{center}
\hspace*{-0.5cm}
\begin{tabular}{c|rrr}  \hline \hline
Integrals & CASCI \hfill & CASCI \hfill & CASCI  \\
          &       & + singles & + singles \\
          &       &           & + diff. doubles \\ \hline
$J_\bot$ & -60.6 & -321.1 & -293.5 \\
$t_\bot$ &-420.6 & -542.7 & -538.2 \\ \hline  \hline 
\end{tabular}
\end{center}

{\normalsize TABLE 2.  $\alpha^\prime N\!aV_2O_5$. Effective exchange
and hopping integrals between the magnetic $d_{xy}$ orbitals of the
two vanadium atoms of the rungs~\cite{vana1}. The calculations have
been performed using the IDDCI$_n$ approach.}

\vspace*{1eM} Comparatively the dynamical correlation effects due to
the double-excitations on the CAS determinants have much smaller
contributions (in general of the order of $10\%$ to
$20\%$~\cite{corrdyn} and much smaller in the present example). I
would like to point out that while the large majority of the
double-excitations on the CAS involve excitations from two core
orbitals towards two virtual orbitals, these excitations do not
significantly contribute to the excitation energies between two
CAS-based states. In fact they exactly cancel out in a second order
perturbative approach. It can therefore be a wise solution to consider
only the differential double-excitations, that is only the
double-excitations that contribute to the energy differences at the
second order of perturbation. This is what is done in the Difference
Dedicated Configuration Interactions method of Malrieu {\it et
al}~\cite{DDCI}.

One should however notice that in the presented results
the bridging orbital have been included in the CAS and that the
crucial bridge polarisation and correlation effects are included in
the single and double-excitations to the CAS. Whether the CAS should
be restricted to the only magnetic orbitals, than triple-excitations
to the CAS should be considered in order to treat these
effects.

\section*{ \underline{\Large 6. From the ab initio wave}\\ 
 [0.7eM]
\underline{\Large function to the model} \\  [0.7eM]
\underline{\Large Hamiltonian}}
\label{sec:heff}

In the simplest cases ---~which are the most numerous~--- the
knowledge of the excitation energies is necessary and sufficient to
determine the integrals amplitudes of the model hamiltonian. This is
for instance the case for the effective exchange integral between
$1/2$ effective spins (as was largely discussed in section~2) or for
the effective hopping between sites, if they are not equivalent by the
space symmetry operations. However things can be sometimes more
complicated. Typically one may encounter two types of difficulties.
\begin{itemize}
\item The problem may be over-determined. \\ This is for instance the
case for an effective spin-1 Heisenberg model. Such models are
particularly adequate for the description of magnetic systems where
the magnetic sites support two unpaired electrons in (quasi-)
degenerate, strongly correlated, orbitals. The Hund's rule imposes a
triplet coupling of these two electrons and, under certain
conditions~\cite{Nicosp1}, they can be modeled by an effective
spin-1. The interaction between two such spin-1 yields a set of
singlet, triplet and quintet states, which wave functions are totally
defined by the spin symmetry, and energy differences by the effective
exchange parameter $J$. One therefore has {\it two} computed target
states excitation energies (singlet to triplet, $\Delta
E_{sg\rightarrow tp}$, and singlet to quintet, $\Delta
E_{sg\rightarrow qt}$) that have no reasons to strictly follow the
spin-1 Heisenberg hamiltonian eigenvalues relationships ($\Delta
E_{sg\rightarrow tp} = J$, $\Delta E_{sg\rightarrow qt} = 3J$), and
only {\it one} parameter ($J$).
\item The problem may be under-determined. \\ This is for instance the
case for a hopping problem between non-equivalent sites~: the
excitation energy  between the two doublet states is not
sufficient to determine both the inter-site hopping and the energy
difference between the two active orbitals.  In such cases, it is
necessary to use the wave functions information.
\item More generally in systems for which the states are not totally
determined by the space or spin symmetry, the energy differences are
not sufficient for the complete and accurate determination of the
effective model. Since the excitation energies cannot provide
the necessary wave function information, it is impossible to meet the
requirement of a good reproduction of the wave-functions physical
content.
\end{itemize}
According to the types of difficulties encountered, several solutions
can be proposed that will be detailed in the next sections.

\subsection*{\underline{\large 6-A. Least square fit}}

In the case of an over-determined problem, one may choose to abandon
the reproduction of some target states and to reduce the objectives of
the model to the accurate description of a smaller set of selected
states. Part of the model hamiltonian eigenstates will thus be present
in the model only as a purpose of completion or in order to allow the
sufficient flexibility to reach the seeked accuracy on the selected
states. A detailed analysis of this option will be given later on.

An alternative choice can be to abandon the exact reproduction of the
ab initio data and rather to model them {\em at the best} using a
least square fit technique. This method has the advantages to describe
equally well (or bad) the whole set of target states. The aim is to
fit, simultaneously and on an equal footing, the computed energies and
target states projections.  One will thus apply the least square fit
on the eigen-equations of the model hamiltonian ($H_{m\!o\!d}$). In a
first approximation the Lagrangian can be defined as
\begin{eqnarray*}
\cal L &=& \sum_i \left| H_{m\!o\!d} P|\Psi_i\rangle - \varepsilon_i
           P|\Psi_i\rangle \right|^2 \\ 
& {\rm or} & {\rm equivalently } \\
\cal L &=& \sum_i \sum_\alpha \left( \langle \Phi_\alpha
           |H_{m\!o\!d}P|\Psi_i\rangle /\langle \Phi_\alpha
           |P|\Psi_i\rangle - \varepsilon_i \right)^2
\end{eqnarray*}  
where $|\Psi_i\rangle$ and $\varepsilon_i$ are respectively the
computed target states wave functions and energies, $P=\sum_\alpha
|\Phi_\alpha\rangle \langle \Phi_\alpha|$ is the active space
projector and $|\Phi_\alpha\rangle$ are the VB configurations
acting as basis set of the active space.  The minimisation of
$\cal L$ runs  on the parameters of $H_{m\!o\!d}$.  The
strict usage of the above formulation presents however some drawbacks.
\begin{itemize}
\item The total energies have no real signification and only the
excitation energies have a physical meaning. One should therefore use
only eigen-equations differences in the Lagrangian.
\item The weight devoted in the Lagrangian to the very small
components of the wave function is much too large compared to the
weight devoted to the large components. Specially since the relative
error made on these small coefficient is much larger and their
physical importance much less. One should therefore ponderate their
contribution in $\cal L$ so that small variations of their amplitude
(but that may be large in relative variations) do not significantly
affect the result.
\end{itemize} 
Finally, as in all least square fit method, the accuracy of the
procedure is measured by amplitude of $\sqrt{\cal L}$ which should be
much smaller than the typical scale of the fitted effective
interactions.

\subsection*{\underline{\large 6-B. Using the wave function information}}

The usage of the computed target states wave functions raises the
question of the reduction of the information contained in a wave
function spanned over several hundred thousands to millions of
determinants onto a model state described by a few VB configurations.

For this purpose it is possible to use the guidelines of the effective
hamiltonian theory~\cite{block}, that is to define the model
hamiltonian by its eigenvalues and eigenvectors.
\begin{itemize}
\item The model hamiltonian eigenvalues ($\epsilon_i$) should be set
to the target states excitation energies computed from the ab-initio
spectroscopy of the embedded fragments ($\varepsilon_i$).
\item The model hamiltonian eigenvectors ($\xi_i$) should be set to the
projection onto the active space of the computed target states wave
functions ($\Psi_i$).
\end{itemize}
These conditions ---~that suppose the equality between the active
space and the target space dimensions ($m_a=m_t$)~--- totally define
the model hamiltonian in a unique manner. However the procedure can be
strictly used only in few special cases since it is often incompatible
with some the of formal requirements demanded from the model.
\begin{itemize}
\item Typically when the target states projections onto the active
space ($P|\Psi_i\rangle$) are not fully defined by the symmetry of the
system, there is no reason for them to form an orthogonal set. The
direct corollary of this non-orthogonality is the lost of the
hermiticity property of the model hamiltonian. Going back to the
example of the effective hopping between two non equivalent sites $a$
and $b$, the two computed target doublets ($|\Psi_1\rangle$ of energy 
$\varepsilon_1$ and $|\Psi_2\rangle$ of energy $\varepsilon_2$)
 can be written as
\begin{eqnarray*}
{P|\Psi_1 \rangle \over \sqrt{\langle \Psi_1|P|\Psi_1 \rangle }}&=& 
\cos{\alpha}\, |a\rangle + \sin{\alpha}\, |b\rangle \\
{P|\Psi_2 \rangle \over \sqrt{\langle \Psi_1|P|\Psi_1 \rangle }}&=& 
-\sin{\beta}\, |a\rangle + \cos{\beta}\, |b\rangle 
\end{eqnarray*}
which yield a model hamiltonian 

\begin{eqnarray*} \hspace*{-1.3cm}
H_{m\!o\!d} &=& \bordermatrix{ 
& |a\rangle & |b \rangle \cr
&{\varepsilon_1 \cos{\alpha}\cos{\beta} + 
 \varepsilon_2 \sin{\alpha} \sin{\beta} \over \cos{(\alpha - \beta)}} &
{(\varepsilon_1 - \varepsilon_2) \cos{\alpha}\sin{\beta}
 \over \cos{(\alpha - \beta)}} \cr && \cr
&{(\varepsilon_1 - \varepsilon_2) \sin{\alpha}\cos{\beta} 
 \over \cos{(\alpha - \beta)}} &
{\varepsilon_1 \sin{\alpha} \sin{\beta} + 
 \varepsilon_2 \cos{\alpha}\cos{\beta} \over \cos{(\alpha - \beta)}}
\cr }
\end{eqnarray*}

One sees immediately that when $\alpha \ne \beta$ the hermiticity is
not verified.
\item In addition, the above procedure yields a priori independent and
non-zero matrix elements between all VB configurations in the same
irreducible representation. In consequence, the number of independent
parameters of the effective model is equal to the number of matrix
elements not directly imposed by the symmetry. The goal of the
modelisation, in terms of the reduction of the physics to a small
number of physically readable, dominant interactions, is therefore
totally lost.
\end{itemize}

The price to retrieve the hermiticity and the simplicity is either (i)
the abandonment of the exact reproduction of the computed excitation
energies and wave functions to the profit of an approximate
representation through a least square fit method (see previous
section), or (ii) the abandonment of the reproduction of part of the
target states in order to focus on the accurate description of a set
of selected states. In the above example of asymmetric hopping the
choices would translate into either the approximate representation of
the two doublets by average states as resulting from a L\"owdin
orthogonalisation procedure
\begin{eqnarray*}
{P|\Psi_1 \rangle \over \sqrt{\langle \Psi_1|P|\Psi_1\rangle }}
&\longrightarrow&
 \cos{\alpha + \beta\over 2} \, |a\rangle +
\sin{\alpha + \beta\over 2} \, |b\rangle \\ 
{P|\Psi_2\rangle  \over \sqrt{\langle\Psi_1|P|\Psi_1 \rangle }}
&\longrightarrow&
 -\sin{\alpha + \beta\over 2} \, |a\rangle +
\cos{\alpha + \beta\over 2} \, |b\rangle
\end{eqnarray*}
or the accurate reproduction of one of the doublets (for instance
$|\Psi_1\rangle$), the second one being accordingly defined for this
purpose (Smidt orthogonalisation)
\begin{eqnarray*}
{P|\Psi_1 \rangle \over \sqrt{\langle \Psi_1|P|\Psi_1\rangle }}&\longrightarrow& 
\cos{\alpha} \, |a\rangle + \sin{\alpha} \, |b\rangle \\ 
{P|\Psi_2 \rangle \over \sqrt{\langle\Psi_1|P|\Psi_1 \rangle }}&\longrightarrow&
 -\sin{\alpha} \, |a\rangle + \cos{\alpha} \, |b\rangle
\end{eqnarray*}

More generally this second choice is related to the guidelines of the
intermediate hamiltonian theory~\cite{hint}~: reproduction by a subset
of the active space of a selected number of target states, both in
terms of energies and wave functions. One should therefore partition
the active space underlying the model hamiltonian into two subspaces. 
\begin{itemize}
\item A main active space of dimension ($m_{ma}<m_a$).  This subspace
will be supported by the eigenstates of the model hamiltonian aimed at
reproducing the target states.  $m_{ma}$ should therefore be equal to
the number of target states to reproduce ($m_{ma}=m_t$).
\item An intermediate active space of dimension $m_{ia} = m_a -
m_{ma}$, supported by the remaining eigenstates of the model
hamiltonian. These states are devoted to provide to the model the
necessary flexibility to both verify the (formal) requirements and to
best reproduce the target states. They have a formal purpose but no
physical purpose even though they may be related ---~however not a
trustworthy representation~--- to meaningful physical states. 
\end{itemize}
Such a procedure, based on a selective choice, send us back to the
aim, in fine, of the model. Indeed the partition of the active space
must be determined by the physics targeted at, that is the states of
the entire lattice one would like to describe. From the embedded
fragment point of view, it means an accurate reproduction of the local
physical content of these infinite system states, that is of the
fragment states which are highly populated (as can be computed by the
density matrices projected on the fragment) in the infinite system
states.

Let us illustrate our purpose by a simple example~: the low energy
physics of a half-filled magnetic system. The infinite system is
therefore a lattice which low energy physics is based on equivalent
magnetic centers supporting each one active orbital and one active
electron. The states responsible for the magnetic properties are
therefore neutral in a Valence-Bond sense, that is
configurations where each magnetic center support a unique active
electron are dominant in the low energy states of the entire system.
Considering now an embedded fragment containing two magnetic centers
(say $A$ and $B$), the active orbitals ($a$ and $b$) and electrons
define a $2\bar e$, $2$ orbitals CAS underlying a Hubbard model
hamiltonian. 
\begin{eqnarray*}
H_{m\!o\!d} &=& \bordermatrix{ 
& |a \bar b\rangle &|b \bar a\rangle & |a \bar a\rangle & |b \bar b\rangle \cr
& 0 & 0 & t & t \cr
& 0 & 0 & t & t \cr
& t & t & U & 0 \cr
& t & t & 0 & U \cr}
\end{eqnarray*}
with the eigenstates
\begin{eqnarray*}
|Sg\rangle &=& \cos\alpha 
 {|a \bar b\rangle + |b \bar a\rangle \over \sqrt{2}} + 
 \sin\alpha 
 {|a \bar a\rangle + |b \bar b\rangle \over \sqrt{2}} \\
&& \text{of energy } {U - \sqrt{U^2+16t^2}\over 2} \\
|Tp\rangle &=& 
 {|a \bar b\rangle - |b \bar a\rangle \over \sqrt{2}} 
 \text{ of energy } 0 \\
|Sg^*\rangle &=& -\sin\alpha 
{|a \bar b\rangle + |b \bar a\rangle \over \sqrt{2}} + 
 \cos\alpha 
 {|a \bar a\rangle + |b \bar b\rangle \over \sqrt{2}} \\
&& \text{of energy } {U + \sqrt{U^2+16t^2}\over 2} \\
|Su\rangle &=& 
 {|a \bar a\rangle - |b \bar b\rangle \over \sqrt{2}} 
 \text{ of energy } U
\end{eqnarray*} 
In a Bloch effective hamiltonian theory the above problem is
over-determined since one has $3$ energy differences and $2$ singlet
eigenstates which wave functions are not totally determined by the
symmetry. The result is $5$ conditions for only $2$ unknown
parameters. The infinite system physics being supported by neutral VB
configurations it is crucial that the local neutral states are best
reproduced. One should therefore prefer an intermediate hamiltonian
guideline and focus on the reproduction of the $|Sg\rangle$ and
$|Tp\rangle$ energy difference and wave functions, yielding a well
defined problem with two conditions (reproduction of the
singlet-triplet energy difference and ionic configurations coefficient
in the singlet state) and two unknowns.  The main consequence of this
choice is that the on-site effective bi-electronic repulsion integral
$U$ is determined so that to best reproduce the correct component in
the singlet ground state of the ionic VB configuration $|a \bar
a\rangle + |b \bar b\rangle$. As a corollary the energies of the
essentially ionic states $|Su\rangle$ and $|Sg^*\rangle$ are
overestimated. Indeed, it can be seen from a simple perturbative
expansion that the dynamical correlation effects on the ionic
configurations strongly depend on whether they belong to an
essentially neutral or essentially ionic state.  In case of a neutral
state the rescaling of the ionic configurations energy by the
dynamical correlation processes is (at the second order of
perturbation)
\begin{eqnarray*}
U &\longrightarrow & U_0 - 
\sum_i {\left|\langle a\bar a |1/r_{12}| \Phi_i\rangle\right|^2 
        \over E_i} 
\end{eqnarray*}
while in case of an ionic state  it comes
\begin{eqnarray*}
U &\longrightarrow & U_0 - 
\sum_i {\left|\langle a\bar a |1/r_{12}| \Phi_i\rangle\right|^2 
        \over E_i-U} 
\end{eqnarray*}
One should thus be conscious that while the model determined under the
previous criteria will be appropriate for the determination of the low
energy physics it is unsuitable to study the transitions 
toward essentially ionic VB states.

In this example the reduction of the number of target states to
reproduce provides a solution to two different problems. The first one
is getting a mathematically well conditioned inversion problem. The
second one is the resolution of the dilemma between an accurate
reproduction of the ground state wave function and of the transition
energy towards ionic excited states. While the second type of problems
can be solved by the usage of intermediate hamiltonian guidelines, the
solution to first type of problems may require such drastic
reduction of the target space (one state per irreducible
representation) that the model may become meaningless. In such a case a
thoughtful combination with square fit techniques will provide the
exit.

\section*{ \underline{\Large 7. Conclusion}}
The fascinating properties observed in the new materials synthesized
in the last couple of decades are commanded by their electronic
structure and more specifically by strong electron-electron
correlations.  The understanding of the observed collective effects
necessitates a clear picture of the interplay between the dominant
microscopic local interactions. Quantum chemists with their extensive
culture of correlation effects in finite systems, of valence
correlation but also dynamical correlation and its retroactive effects
on the valence shell, in the ground state but also in excited,
multi-reference states, have the intellectual tools to answer this
question. We have seen in this paper how up-to-date quantum chemical
spectroscopic methods associated to a controlled description of the
crystalline embedding, a clear understanding of the model hamiltonians
aims, an appropriate extension and usage of effective hamiltonian
methods that allow a controlled reduction of the ab initio information
into a few dominant interactions, are the clues to a controlled and
accurate modelisation of these materials. 

It is a fascinating challenge for the quantum chemist used to small
finite systems, to  enter this condensed matter field where the
answers can only be find by the integration over all physical scales
from first principle microscopic local electronic interactions, to the
nanoscopic scale (fragments) up to the macroscopic properties.

\newpage
\section*{ \underline{\Large Appendix}}

\subsection*{ \underline{\large The Heisenberg model} }
The Heisenberg model was first devised by Heisenberg~\cite{heis1} and
formalized by Van Vleck and Dirac~\cite{heis2} in order to rationalize
isotropic ferromagnetic interactions between localized electrons or
spin $1/2$ particles supported by orbitals localized on well separated
centers. Its validity has later been extended by Anderson~\cite{and}
to anti-ferromagnetic interactions by the famous super-exchange
mechanism~: stabilisation of the singlet state compared to the triplet
state by the effective effect of the ionic configurations.  The
general expression of the Heisenberg Hamiltonian is for a system with
one spin, one orbital per center (half-filling)
\begin{eqnarray}
\label{eq:heis1}
H_{Heis}&=& \sum_{<i,j>} J_{ij} \left({\bf S}_i . {\bf S}_j -
{1\over 4}n_in_j\right) 
\end{eqnarray}
When the spins represent electrons (spin $1/2$) the Heisenberg Hamiltonian can be rewritten as 
\begin{eqnarray} \hspace*{-0.7cm}
\label{eq:heis2}
H_{Heis}&=& \sum_{<i,j>} J_{ij} 
\left({ a^\dagger_{i\uparrow}a^\dagger_{j\downarrow} - a^\dagger_{j\uparrow}a^\dagger_{i\downarrow}  \over \sqrt{2}} \right)
\left( { a_{i\uparrow}a_{j\downarrow} - a_{j\uparrow}a_{i\downarrow}  \over \sqrt{2}} \right)
\end{eqnarray}
where $a^\dagger_{i\sigma}$ (resp. $a_{i\sigma}$) is the creation
(resp. annihilation) operator of an electron of spin $\sigma$ on site
$i$, $n_i$ is the number operator on site $i$ and ${\bf S}_i$ the spin
operator on site $i$. The sum over $<i,j>$ runs on nearest neighbors
sites $i$ and $j$.

\subsection*{ \underline{\large The $t-J$ model} }

The $t-J$ model was first devised by Zhang and Rice~\cite{t-J} in
order to described the movement of holes in an anti-ferromagnetic
background such as found in the copper oxide planes of the cuprate high
$T_c$ super-conductors. It applies to magnetic systems away from
half-filling (number of magnetic orbitals different from the number of
electrons or spins). According to the sign of the doping (extra
electrons or extra holes compared to the half-filled system) the
hamiltonian applies to the electrons or holes. Its general form is
\begin{eqnarray}
\label{eq:tJ}
H_{t-J}&=& \sum_{<i,j>} J_{ij} \left({\bf S}_i . {\bf S}_j 
 -{1\over 4}n_in_j\right) + \nonumber \\ && 
\sum_{<i,j>} t_{ij} \sum_\sigma \left(a^\dagger_{i\sigma}a_{j\sigma} + 
a^\dagger_{j\sigma}a_{i\sigma}\right) 
\end{eqnarray}

\subsection*{\underline{\large The extended Hubbard model}}

The extended Hubbard model is basically a simplified version of the
Pariser-Parr-Pople hamiltonian. It contains essentially four types of
interaction parameters, two mono-electronic ones corresponding to the
H\"uckel approximation and two bielectronic ones
\begin{itemize}
\item the orbitals energies, 
\item the hopping integrals (usually limited to adjacent centers),
\item the bielectronic coulomb repulsion between two electrons located
on the same orbital,
\item and the bielectronic coulomb repulsion between two electrons
located on different orbitals (most of the time limited to nearest
neighbors centers).
\end{itemize}
The three first terms define the Hubbard model and the addition of the
inter-site repulsions leads to the extended Hubbard model. It comes the following generic formulation
\begin{eqnarray}
\label{eq:hub}
H_{Hub} &=& 
\sum_{<ij>} t_{ij} \sum_\sigma \left(a^\dagger_{i\sigma} a_{j\sigma} + a^\dagger_{j\sigma} a_{i\sigma}\right) + \nonumber \\ &&
\sum_i U_i n_{i\uparrow}n_{i\downarrow} + \\&&
\sum_{<ij>} V_{ij}\left(n_i-Z_i\right)\left(n_j-Z_j\right) \nonumber
\end{eqnarray} 
where $Z_j$ is the effective nucleus charge of the center $i$.

\end{document}